\documentclass[fleqn,10pt]{wlscirep}
\usepackage{jabbrv}
\usepackage[utf8]{inputenc}
\usepackage[T1]{fontenc}
\usepackage{bm}
\usepackage{siunitx}
\usepackage{amssymb}
\title{Reconfigurable Digital RRAM Logic Enables In-situ Pruning and Learning for Edge AI}

\author[1,2,3$^{\dagger}$]{Songqi Wang}
\author[1,2$^{\dagger}$]{Yue Zhang}
\author[4,5]{Jia Chen}
\author[1,3]{Xinyuan Zhang}
\author[1]{Yi Li}
\author[1]{Ning Lin}
\author[1]{Yangu He}
\author[1,3]{Jichang Yang}
\author[5]{Yingjie Yu}
\author[5]{Yi Li}
\author[1,2,4*]{Zhongrui Wang}
\author[1*]{Xiaojuan Qi}
\author[1,3*]{Han Wang}

\affil[1]{Department of Electrical and Electronic Engineering, the University of Hong Kong, Hong Kong, China}
\affil[2]{School of Microelectronics, Southern University of Science and Technology, Shenzhen, China.}
\affil[3]{Center for Advanced Semiconductor and Integrated Circuit, The University of Hong Kong, Hong Kong, China}
\affil[4]{ACCESS – AI Chip Center for Emerging Smart Systems, InnoHK Centers, Hong Kong Science Park, Hong Kong, China}
\affil[5]{School of Optical and Electronic Information
Huazhong University of Science and Technology, China}

\affil[$^{\dagger}$]{These authors contributed equally.}
\affil[*]{e-mail: wangzr@sustech.edu.cn; xjqi@eee.hku.hk;
hanwang6@hku.hk}

\begin{abstract}
The human brain simultaneously optimizes synaptic weights and topology by growing, pruning, and strengthening synapses while performing all computation entirely in memory.
In contrast, modern artificial‑intelligence systems separate weight optimization from topology optimization and depend on energy‑intensive von Neumann architectures. Here, we present a software–hardware co‑design that bridges this gap.
On the algorithmic side, we introduce a real‑time dynamic weight‑pruning strategy that monitors weight similarity during training and removes redundancies on the fly, reducing operations by 26.80\% on MNIST and 59.94\% on ModelNet10 without sacrificing accuracy (91.44\% and 77.75\%, respectively).
On the hardware side, we fabricate a reconfigurable, fully digital compute‑in‑memory (CIM) chip based on 180 nm one-transistor-one-resistor (1T1R) RRAM arrays. Each array embeds flexible Boolean logic (NAND, AND, XOR, OR), enabling both convolution and similarity evaluation inside memory and eliminating all ADC/DAC overhead. The digital design achieves zero bit-error, reduces silicon area by 72.30\% and overall energy by 57.26\% compared to analogue RRAM CIM, and lowers energy by 75.61\% and 86.53\% on MNIST and ModelNet10, respectively, relative to an NVIDIA RTX 4090.
Together, our co‑design establishes a scalable brain‑inspired paradigm for adaptive, energy‑efficient edge intelligence in the future.

\end{abstract}

\begin{document}
\flushbottom
\maketitle
\thispagestyle{empty}

\section*{Introduction}
\begin{figure}[!t]
\centering
\includegraphics[width=0.9\linewidth]{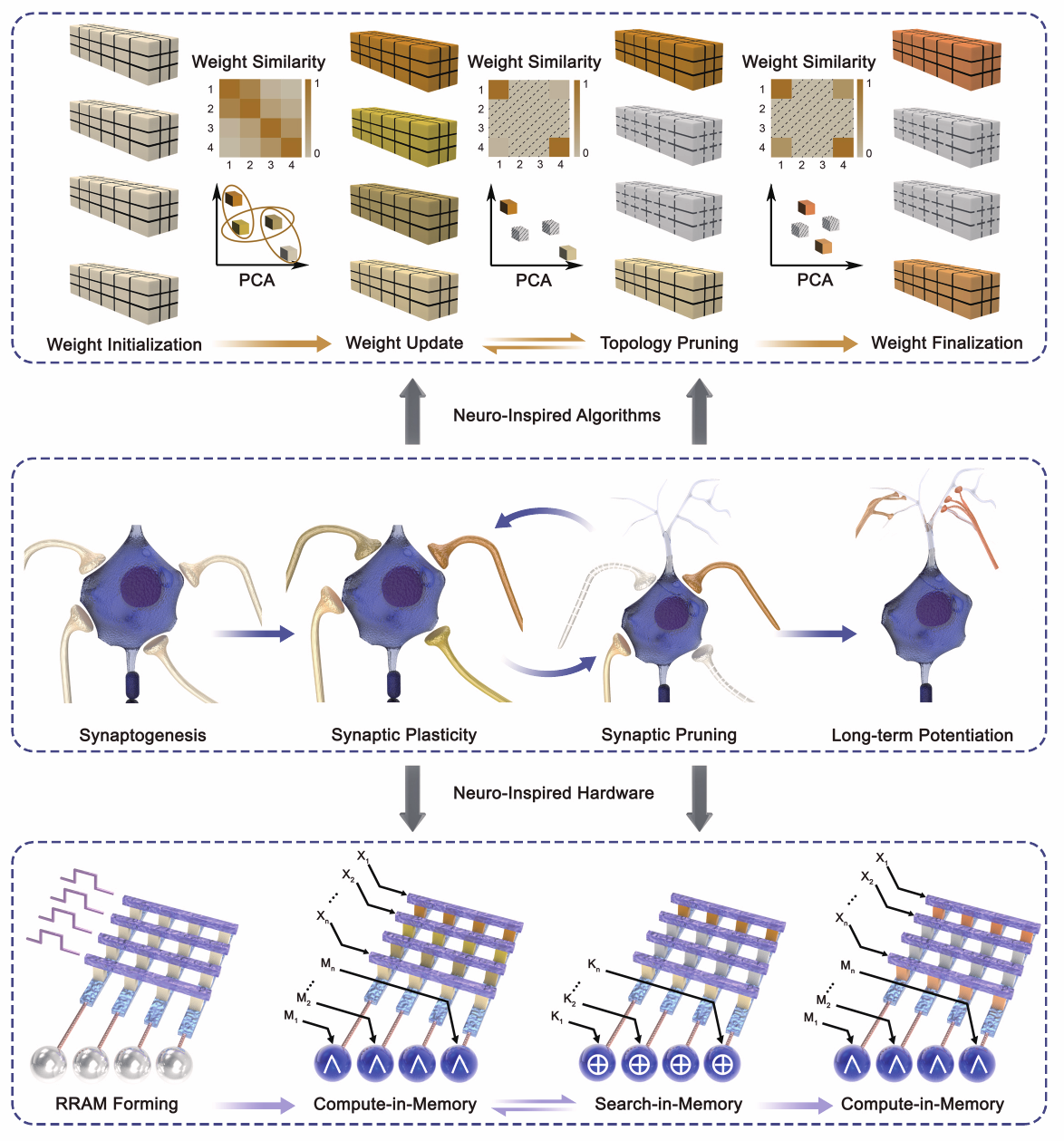}
\caption{\textbf{Neuro‑inspired algorithm–hardware co‑design for simultaneous weight and topology optimization.}
\textbf{a,} Software pipeline that iteratively performs \emph{Weight Initialization}, followed by alternating cycles of \emph{Weight Update} and \emph{Topology Pruning}. Weight‑similarity heat maps and principal‑component clusters reveal redundant weights, which are removed on the fly; the retained weights are subsequently fine‑tuned to recover accuracy. After multiple iterations, weights and topology converged in the \emph{Weight Finalization} stage. \textbf{b,} Biological mechanisms including \emph{synaptogenesis} (formation of synaptic connections), alternating cycles of \emph{synaptic plasticity} (adaptive responses to stimuli) and \emph{synaptic pruning} (removal of redundant connections), ultimately leading to \emph{long-term potentiation}, provide the conceptual blueprint for the algorithm described in \textbf{a} and the hardware implementation presented in \textbf{c}. \textbf{c,} Neuro-inspired hardware implementation: Workflow of the fully digital, reconfigurable RRAM CIM chip, illustrating the stages of RRAM array electroforming (initialization), alternating compute-in-memory (convolution computation) and search-in-memory (logic operations for weight similarity detection).
}
\label{fig1}
\end{figure}

The human brain exhibits two intertwined hallmarks:  
(i) lifelong co‑optimization of synaptic weights and network topology; and  
(ii) exceptional energy efficiency~\cite{faust2021mechanisms,watanabe2011climbing} (Fig.~\ref{fig1}b).
Hallmark (i) is sustained by continuous \emph{synaptogenesis} together with alternating cycles of \emph{synaptic plasticity} and \emph{synaptic pruning}, which together dynamically adjust neural circuit connectivity throughout life, ultimately leading to \emph{long-term potentiation}~\cite{andjus2003change,hashimoto2011postsynaptic}. In the early phases of learning, synaptogenesis diversifies connections, expanding the network to accommodate new stimuli and increase storage capacity~\cite{miyazaki2012cav2,mikuni2013arc}.  
As activity evolves, synaptic plasticity strengthens frequently co‑activated synapses while weakening those seldom used, via Hebbian and homeostatic rules~\cite{lorenzetto2009genetic,kawamura2013spike}.  
To curb complexity, microglia‑mediated synaptic pruning removes redundant or inefficient links, preserving only the most informative pathways~\cite{nakayama2012gabaergic,kano1995impaired,paolicelli2011synaptic,ichikawa2016territories,depression1999mglur1}.  
These simultaneous synaptic strength and topology optimizations iterate in a lifelong manner.
With respect to hallmark (ii), information is stored and processed \emph{in situ} at synapses; consequently, neural computation minimises data movement and thus power consumption—constituting a natural realisation of compute‑in‑memory (CIM) principles~\cite{strukov2008missing,rao2023thousands,wan2022compute,ambrogio2018equivalent,le202364,zhang2023edge,rasch2024fast,mutlu2022modern, chen2025refreshable,cheong2024stochastic,dang2024reconfigurable,feng2024memristor}.

Although modern AI systems aspire to emulate the brain, they remain fundamentally different. From a software perspective, conventional deep learning models are not designed to simultaneously co‑optimize weights and network topology over a lifetime. Instead, they typically employ a two-phase approach: full-model weight learning followed by post-hoc topology optimization (e.g. pruning). This bifurcated workflow entails two key drawbacks: (i)  The maintenance of a fixed parameter set throughout training incurs excessive computational cost and energy consumption~\cite{lin2020dynamic}, as many weights are redundant and contribute little to inference accuracy~\cite{lu2024redtest, liu2017kernel}. (ii)  Post‑training pruning often leads to accuracy loss due to insufficient parameter adaptation~\cite{wang2023ntk}. These limitations are especially detrimental to resource‑constrained edge devices, where both storage capacity and computational power are limited~\cite{park2023dynamic, huang2024fedmef, lu2019super, liu2018dynamic, chen2019drop, elkerdawy2022fire}.

From a hardware perspective, different from the brain's colocation of memory and processing, mainstream digital AI platforms separate memory and processing units, adhering to the traditional von Neumann architecture~\cite{yu2021compute,jain2025heterogeneous,wang2024memristor,yi2023activity,yu2025full}. This physical separation leads to frequent data movement, significantly increasing energy consumption and computational latency. For example, accessing off-chip DRAM consumes up to 1,000 times more energy than performing a 32-bit floating-point multiplication~\cite{horowitz20141, jeong2025self, ning2023memory, fuller2019parallel}. This issue becomes particularly severe in edge devices, where low power consumption and high efficiency are essential~\cite{huo2022computing, khan2025video, zhang2020artificial}. Emerging CIM architectures aim to overcome this bottleneck by integrating matrix operations directly within memory arrays~\cite{roy2024energy, yang2013memristive, liu2025memristor, wolters2024memory, li2018analogue}, thereby reducing data transfer and improving energy efficiency~\cite{li2018efficient, yuan202514, khwa202514}. However, most analog CIM implementations suffer from three limitations. (i) They only support vector-matrix multiplication without reconfigurable logic operations, such as those required for simultaneous weight and topology optimization~\cite{liu2022reconfigurable, pei2025ultra}. (ii) Analog signal noise due to process variations and parasitic resistance compromises analog computing precision~\cite{haensch2022co, wang2020resistive, shi2025memristor}. (iii) Peripheral ADCs and DACs needed to interface with digital systems add substantial energy and area overhead, diminishing the overall benefits of CIM~\cite{lanza2025growing, wang2017memristors, wang202514}.

To address the software limitations, we propose a real-time dynamic weight pruning algorithm inspired by the brain’s ability to simultaneously co-optimize weights and topology. The method monitors inter‑weight similarity during training and excises functionally redundant weights, thereby (i) markedly reducing computational workload and (ii) maintaining predictive performance. As illustrated in Fig.~\ref{fig1}a, the pipeline comprises \emph{Weight Initialization} followed by alternating cycles of \emph{Weight Update} and \emph{Topology Pruning}, culminating in \emph{Weight Finalization}; this computational pipeline mirrors the biological processes of \emph{Synaptogenesis}, \emph{Synaptic Plasticity}, \emph{Synaptic Pruning}, and \emph{Long-Term Potentiation}. During weight initialization, weights are randomly initialized. Weight Update refines these weights through backpropagation. Real-time weight similarity monitoring, visualized through heatmaps and PCA-based clustering, identifies redundant weights for pruning. This dynamic pruning process reduces computational redundancy and optimizes the network structure while preserving key feature extraction capabilities. This simultaneous weight and topology optimization cycle iterates, optimizing the remaining weights and ensuring efficient and accurate inference even in resource-constrained environments. On the MNIST~\cite{lecun1998mnist} and ModelNet10~\cite{wu20153d} datasets, our method reduces operations by 26.80\% and 59.94\%, respectively, while retaining classification accuracies of 91.40\% and 77.75\%.

On the hardware side, we propose a fully digital RRAM-based CIM chip that eliminates the von Neumann bottleneck. Its advantages are threefold. (i) The chip features reconfigurable logic units supporting NAND, AND, XOR, and OR operations~\cite{yan20221}, allowing simultaneous weight and topology optimization. (ii) All computations are performed in the digital domain, thereby eliminating analogue noise and its attendant accuracy loss. (iii) The design avoids frequent analog-to-digital and digital-to-analog conversions, reducing power consumption and simplifying the design complexity. In addition, leveraging the non-volatile nature and high storage density of RRAM, the chip ensures stable weight storage while substantially reducing the overall chip area, making it particularly suitable for energy‑constrained edge computing applications. As illustrated in Fig.~\ref{fig1}c, the process begins with weight initialization, where RRAM cells are initialized to stable, random resistance states through forming voltage pulses. During the CIM stage, logical operations are executed directly within the memory array and peripheral circuits. Quantized input is encoded as high and low voltage levels to perform AND operations. In the search-in-memory stage, XOR operations compute Hamming distances or Euclidean distances to facilitate weight similarity analysis for pruning. Pruned weights are represented as gray RRAM cells, indicating deactivation to reduce computational overhead. As the alternative CIM and search-in-memory stages iterate, the memory array and peripheral circuits continue learning, thereby ensuring that the pruned neural network maintains high inference accuracy and reliable performance under resource-constrained conditions.

\begin{figure}[!t]
\centering
\includegraphics[width=0.9\linewidth]{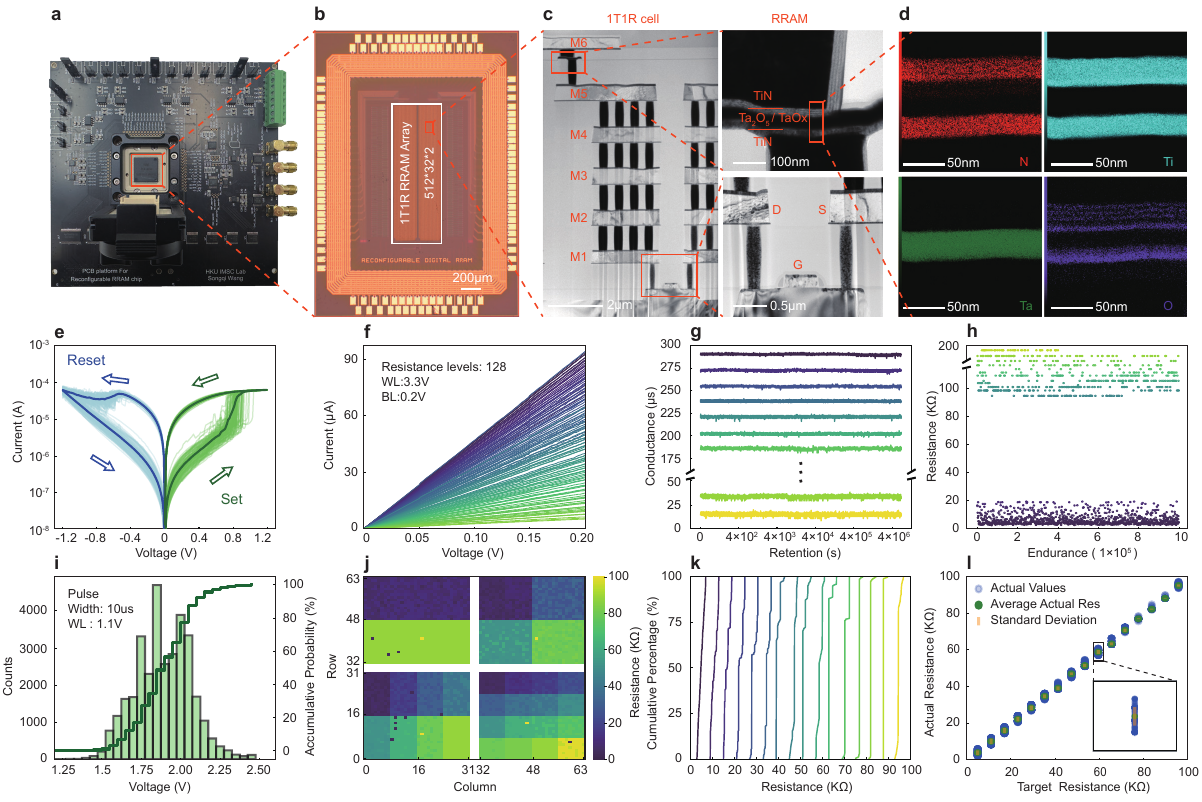}
\caption{\textbf{Physical and Electrical Characterization of the RRAM array.}
\textbf{a,} Optical image of the system board. The RRAM-based reconfigurable logic chip is accommodated by the center socket.
\textbf{b,} Magnified optical image of the reconfigurable logic chip, showing \(512\times32\times2\) RRAM array and peripheral circuit.
\textbf{c,} Cross-sectional transmission electron microscope (TEM) images of the 1-Transistor-1-RRAM (1T1R) cell, illustrating the NMOS transistor and RRAM material stack.
\textbf{d,} Energy-dispersive X-ray spectroscopy (EDS) mapping of Ti, N, O, and Ta in an RRAM cell.
\textbf{e,} Quasi-static I-V characteristics showing bipolar resistive switching of the RRAM.
\textbf{f,} Multilevel programming with 128 distinct resistance states.
\textbf{g,} Retention test of RRAMs in different conductances over $4 \times 10^6$ seconds.
\textbf{h,} Endurance reliability exceeding $1 \times 10^6$ switching cycles.
\textbf{i,} Histogram of electroforming voltages of the entire RRAM array.
\textbf{j,} Evaluation of multi-level programming accuracy by programming the RRAM array to 2, 4, 8, and 16 resistance states.
\textbf{k,} Distribution of RRAM array resistance programmed into 16 distinct resistance levels in j.
\textbf{l,} Measured vs. target resistance values.
}  
\label{fig2}
\end{figure}

\section*{Physical and Electrical Characteristics of RRAM Chip}

This section provides a comprehensive analysis of the physical and electrical characteristics of individual RRAM devices and their array-level performance (see Supplementary Fig.1 for test system setup).

Fig.~\ref{fig2}a illustrates the system board. The optical micrograph in Fig.~\ref{fig2}b presents the unpackaged RRAM chip fabricated in a 180 nm CMOS process, with a total area of 5.016 \text{mm}$^2$. The chip consists of word-line (WL), source-line (SL), and bit-line (BL) driver modules, two \(512\times32\) RRAM arrays, and a reconfigurable logic computing module. Fig.~\ref{fig2}c shows a cross-sectional transmission electron microscope (TEM) image of a one-Transistor–one-RRAM (1T1R) RRAM cell, which is positioned between the metal interconnect layers M5 and M6. The device adopts a TiN/TaO\textsubscript{x}/Ta\textsubscript{2}O\textsubscript{5}/TiN sandwich structure, where the formation and rupture of conductive filaments within the Ta\textsubscript{2}O\textsubscript{5} layer serve as the primary resistive switching medium. Elemental mapping through energy-dispersive X-ray spectroscopy (Fig.~\ref{fig2}d) reveals the spatial distribution of Ti, N, O, and Ta, which is consistent with the RRAM cell design. 

Next, we examine the electrical performance of a single RRAM cell and array. The single cell I-V characteristics (Fig.~\ref{fig2}e) exhibit repeatable bipolar resistive switching, with a set voltage between approximately +0.8 to +0.9 V and a reset voltage between approximately –0.7 to –1.0 V. Multi-level programming capability (Fig.~\ref{fig2}f) is verified by iterative programming, achieving 128 distinct resistance states under a read voltage of 0.3 V (see Supplementary Fig.2 about the programming method). Retention characteristics (Fig.~\ref{fig2}g) indicate that both high-conductance and low-conductance states remain stable for up to \(4\times10^6\) s under a read voltage of 0.3 V at room temperature. No significant conductance drift is observed during the retention test period. Endurance testing (Fig.~\ref{fig2}h) confirms that the device can withstand over \(10^6\) switching cycles while maintaining a stable resistance window. At the array level, the electroforming voltage (\(V_{\text{form}}\)) distribution (Fig.~\ref{fig2}i) follows a normal distribution, with a mean voltage of 1.89 V and a standard deviation of 0.18 V. All devices exhibit successful forming under the applied conditions, resulting in a 100\% forming yield. The programming accuracy of a 32$\times$32 sub-array (Fig.~\ref{fig2}j) is evaluated across 2, 4, 8, and 16 resistance levels, with 99.8\% of devices successfully programmed within a fixed tolerance window of $\pm$2\,k$\Omega$ around each target resistance level. The statistical distribution of resistance values for 16 programmed states (Fig.~\ref{fig2}k) further confirms analog and accurate resistance tuning across the array. Fig.~\ref{fig2}l compares target and actual resistance values, revealing a mean programming standard deviation of $0.8793~k\Omega$.

\begin{figure}[!t]
\centering
\includegraphics[width=0.9\linewidth]{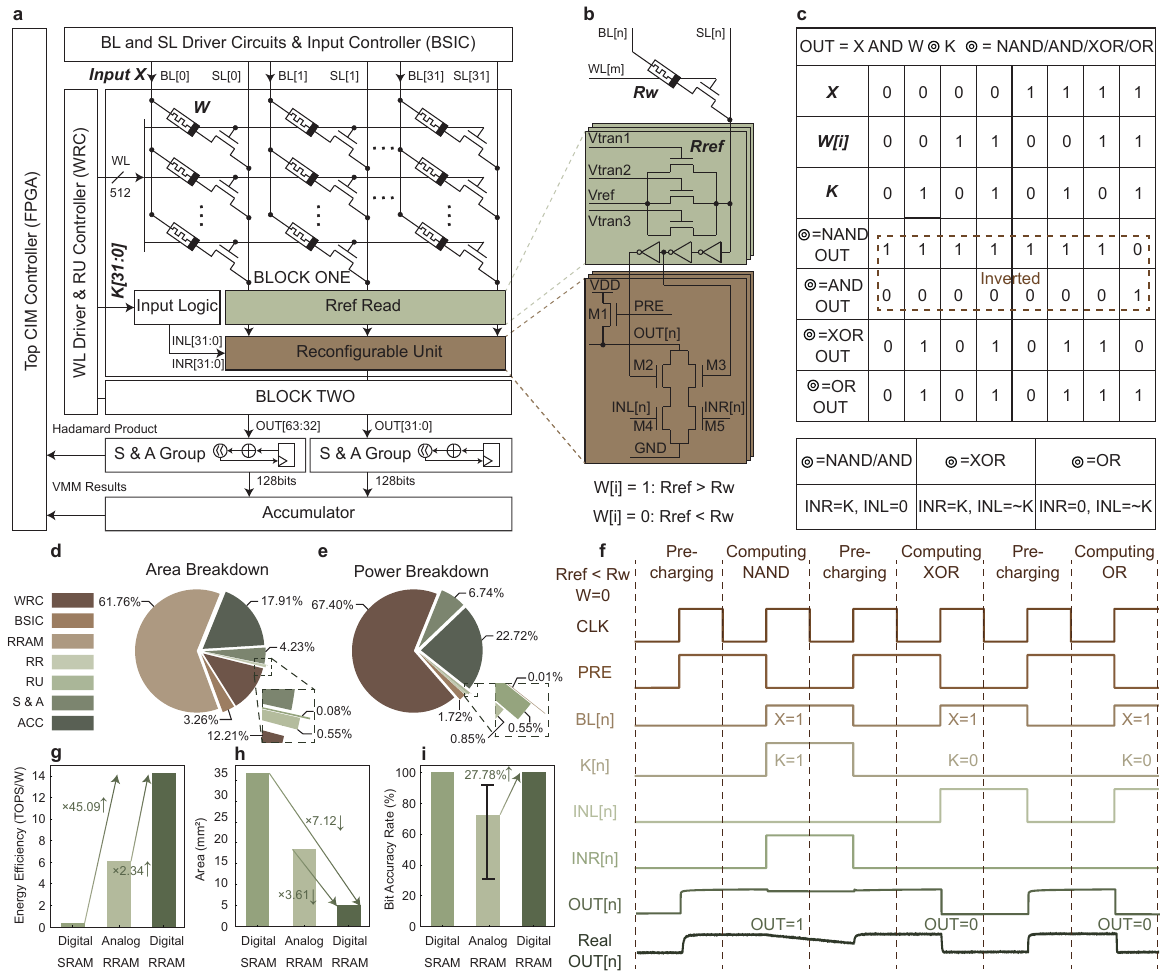}
\caption{\textbf{Chip Architecture and Performance Analysis.}  
\textbf{a,} System-level block diagram of the RRAM-based reconfigurable logic chip. The system comprises a Top CIM Controller, BL and SL Driver Circuits $\&$ Input Controller (BSIC), WL Driver $\&$ RU Controller (WRC), two \(512\times32\) RRAM arrays, Rref Read module (RR), Reconfigurable Unit (RU), Shift and Adder (S$\&$A) groups, and Accumulator (ACC).  
\textbf{b,} Circuit-level implementation of the RU and RR module.
\textbf{c,} Truth table of the ternary logic operations performed by the chip, where \( OUT = X\) and \(W \circledcirc K \). Here, \( X \) represents the input from the BL, \( W[i] \) corresponds to the weight stored in the RRAM cell, and \( K \) is the input to the RU module. The lower table defines the INR and INL values for different logic operations.  
\textbf{d,} Chip area breakdown, where the three largest components are RRAM, ACC, and WRC.  
\textbf{e,} Power consumption breakdown, with WRC, ACC, and S$\&$A contributing the highest energy consumption, while the RRAM array accounts for only \( 0.01\% \).  
\textbf{f,} Timing diagram illustrating the pre-charging and computing phases during NAND, XOR, and OR operations.  
\textbf{g, h, i,} Comparison of energy consumption, chip area, and bit accuracy among digital SRAM-based CIM, analog RRAM-based CIM, and the proposed digital RRAM-based CIM system.
}
\label{fig3}
\end{figure}

\section*{Chip Architecture and Performance Evaluation}

This section provides a detailed analysis of the chip architecture, logic operations, and performance evaluation in comparison to other CIM architectures.

As illustrated in Fig.~\ref{fig3}a, the system consists of a Top CIM Controller implemented on an FPGA (ZCU102) for data exchange and control signal management. The BL and SL Driver Circuits \& Input Controller (BSIC) include decoders that select a specific bit line (BL) during RRAM programming or transmit input signals to all bit lines during computation. The WL Driver \& RU Controller (WRC) employs shift registers to select the corresponding word lines (WL) during both RRAM programming and computation. The core computing unit comprises two \(512\times32\) 1T1R RRAM arrays, designated as Block One and Block Two. Each RRAM cell supports multi-bit storage and achieves a zero bit error rate (BER) for 2-bit storage in our experiments. The Rref Read (RR) module is a resistive divider readout circuit, comparing RRAM resistance values with tunable reference resistance and returning a logic 0 or 1. Another input signal, K, is processed by the Input Logic module, which adjusts control signals based on the required logic operation before transmitting INR and INL signals (as defined in Fig.~\ref{fig3}c) to the Reconfigurable Unit (RU). For element-wise Hadamard product operations, only the S \& A Group is activated, while vector-matrix multiplication (VMM) additionally activates the Accumulator module to sum partial products. Fig.~\ref{fig3}b details the RR and RU circuit implementation. Each 1T1R RRAM cell is connected to the source line (SL) and a tunable reference resistor (Rref), which is adjusted via three NMOS transistors (Vtran1, Vtran2, and Vtran3). The resulting voltage division, representing the resistance comparison outcome, is processed through three inverters before being input into the Reconfigurable Unit (RU), which consists of five NMOS transistors. The timing diagram in Fig.~\ref{fig3}f illustrates the execution of NAND, XOR, and OR operations in both the pre-charging and computing phases (see Supplementary Fig. 4 for the analysis of the leakage path). The truth table in Fig.~\ref{fig3}c defines the logical operations, formulated as: OUT = X AND W $\circledcirc$ K where $\circledcirc$ represents NAND, AND, XOR, or OR operations. The lower table in Fig.~\ref{fig3}c specifies INR and INL values under different logic configurations. 

The chip’s footprint and power breakdown are analyzed in Fig.~\ref{fig3}d and Fig.~\ref{fig3}e. The RRAM array occupies the largest area (61.76\%), followed by the Accumulator (17.91\%) and WRC (12.21\%). In terms of power consumption, the WRC (67.40\%), ACC (22.72\%), and S \& A Group (6.74\%) contribute the highest power usage, while the RRAM array consumes only 0.01\% (see Supplementary table 1 for exact values). Finally, Fig.~\ref{fig3}g,h,i present a comparative analysis of the proposed digital RRAM-based CIM, digital SRAM-based CIM, and analog RRAM-based CIM under identical process technology and storage capacity conditions. The proposed architecture demonstrates a 45.09× improvement in energy efficiency over SRAM-based CIM and a 2.34× improvement compared to analog RRAM-based CIM due to DAC and ADC-free design. In terms of area efficiency, the proposed design achieves a 7.12× reduction in chip area relative to SRAM-based CIM and a 3.61× reduction compared to analog RRAM-based CIM due to DAC and ADC  free design. Furthermore, while digital SRAM-based CIM maintains 100\% bit accuracy, analog RRAM-based CIM exhibits a significantly higher average error rate of 27.78\% (depending on the degree of parallelism) due to intrinsic programming stochasticity. By leveraging digital RRAM-based computation and integrated error correction mechanisms, the proposed system successfully achieves 100\% accuracy.

\begin{figure}[!t]
\centering
\includegraphics[width=0.9\linewidth]{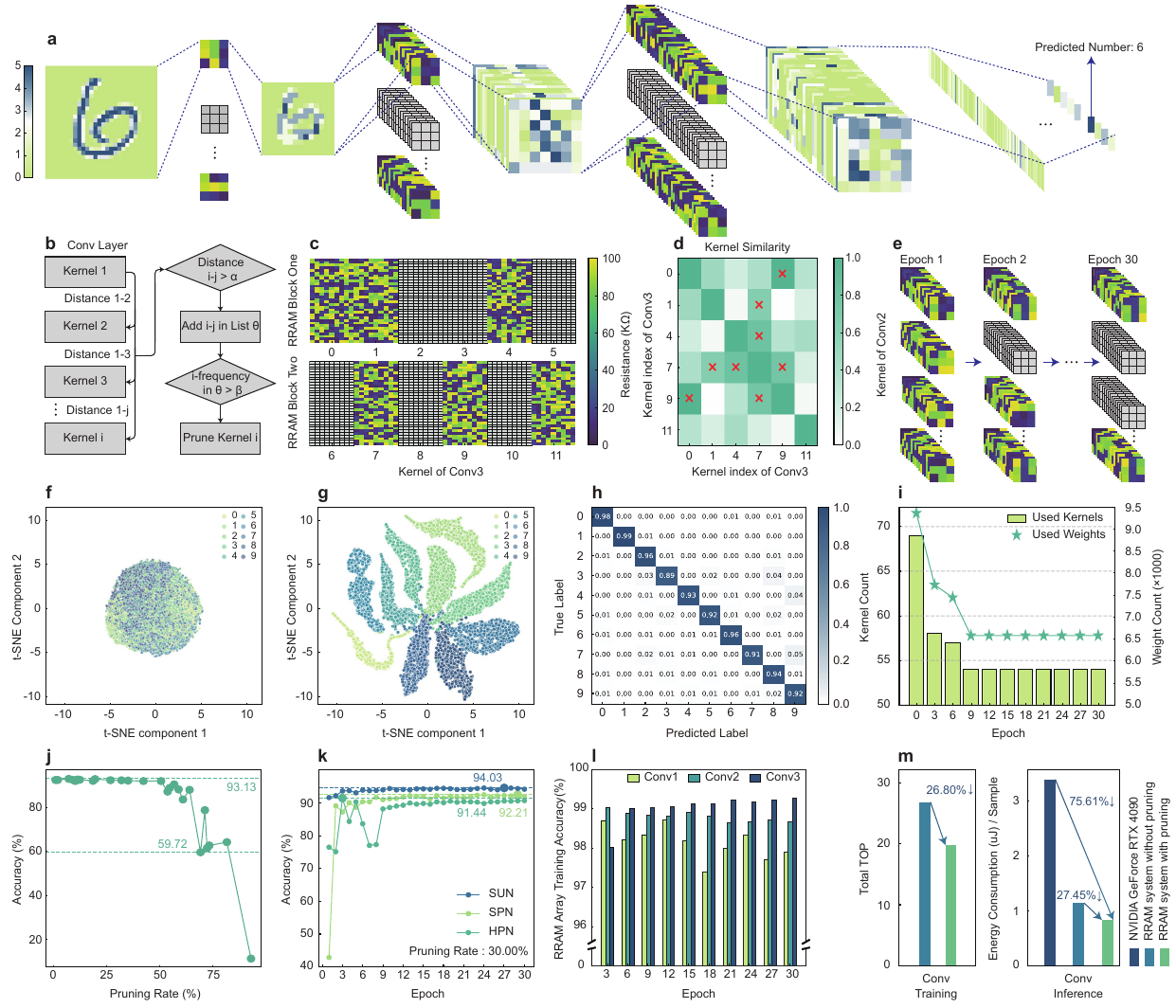}
\caption{\textbf{Dynamic CNN kernel pruning for MNIST classification.}  
\textbf{a,} Schematic illustration of CNN forward pass with an MNIST image "6", illustrating hierarchical feature extraction by convolutional layers and prediction by fully connected layers. Pruned kernels are shown in gray. 
\textbf{b,} Flowchart of the dynamic pruning algorithm, which identifies and removes redundant kernels based on similarity analysis and frequency thresholds.  
\textbf{c,} Mapping of convolutional layer 3 (Conv3) kernels onto the RRAM array.  
\textbf{d,} Kernel similarity matrix of \textbf{c} quantifies redundancy among kernels, where red crosses indicate high similarity. 
\textbf{e,} Epoch-wise kernel pruning, demonstrating the gradual reduction in active kernels during training.  
\textbf{f-g,} t-SNE visualization of the feature space before training and after training with dynamic kernel pruning.
\textbf{h,} Confusion matrix illustrating classification accuracy after pruning.  
\textbf{i,} Reduction in active kernels and total weights over training epochs.
\textbf{j,} Classification accuracy as a function of pruning rate.  
\textbf{k,} Comparison of training accuracy among software-unpruned network (SUN), software-pruned network (SPN), and hardware-pruned network (HPN).  
\textbf{l,} Layer-wise evaluation of MAC precision accuracy for Conv1, Conv2, and Conv3 on the RRAM array, reflecting small hardware variations.  
\textbf{m,} Left: Tera operations (TOP) of Conv layers during training. Right: energy consumption comparison of Conv layers and fully connected (FC) layers during inference between the NVIDIA GeForce RTX 4090 and the digital RRAM system with and without pruning.}
\label{fig4}
\end{figure}

\section*{Dynamic CNN Kernel Pruning for MNIST Classification}

This section presents the implementation and evaluation of dynamic kernel pruning an RRAM-based convolutional neural network (CNN), tested on MNIST classification. The key innovation lies in leveraging the same set of weights for multiple logic operations, where XOR computes kernel similarity for pruning while AND performs forward propagation in the neural network. By jointly learning the kernels to be pruned and the neural network weights, the proposed system significantly reduces training and inference costs while maintaining classification accuracy.

As shown in Fig.~\ref{fig4}a, the CNN consists of three convolutional layers followed by a fully connected layer. The input is a 28×28 grayscale image, progressively processed through convolution, pooling, and activation layers before being flattened and passed to the fully connected layer for classification into ten categories (see Supplementary Table 2 for details). Certain convolutional kernels across all layers are dynamically pruned during training based on kernel similarity. Fig.~\ref{fig4}e illustrates the progressive pruning process over 30 training epochs. The corresponding method, detailed in Fig.~\ref{fig4}b, involves three sequential steps. First, the Hamming distance between convolutional kernels across the entire network is calculated; kernel pairs with distances exceeding a predefined threshold are included in a candidate list. Second, the frequency with which each kernel appears in this list is assessed. Finally, kernels surpassing a preset frequency threshold are pruned. Fig.~\ref{fig4}c presents the mapping of 12 kernels from the third convolutional layer onto the RRAM array at epoch 20. Among them, six kernels are retained in epoch 20, and the next kernel 7 is pruned according to the experimentally measured similarity as shown in Fig.~\ref{fig4}d, where red crosses indicate excessive similarity.

To assess the pruning effectiveness, t-SNE dimensionality reduction is applied to the final-layer outputs. Feature clustering before training is visualized in Fig.~\ref{fig4}f, while Fig.~\ref{fig4}g shows the improved feature separability after RRAM-based training with dynamic kernel pruning (see Supplementary Fig.5 about unpruning network feature separation). Classification performance after pruning is analyzed using the normalized confusion matrix in Fig.~\ref{fig4}h. The x-axis represents predicted labels, the y-axis represents true labels, and the color intensity indicates prediction confidence (dominated by the diagonal line). Fig.~\ref{fig4}i depicts the evolution of active kernels and total weights across training epochs. During epochs 0–3, kernel and associated weight counts drop significantly, reflecting the early-stage removal of redundant kernels. Between epochs 6–30, the pruning rate (or kernel count) stabilizes, maintaining an efficient, compact model structure. Fig.~\ref{fig4}j shows the simulated impact of different pruning rates on classification accuracy. For pruning rates below 50.00\%, accuracy remains stable at 93.13\%. However, once the pruning rate exceeds 50.00\%, accuracy begins to decline rapidly, reflecting the loss of critical feature representations. Fig.~\ref{fig4}k compares simulated training accuracy over epochs for software-unpruned, software-pruned, and hardware-pruned networks under a 30.00\% pruning rate. The software-unpruned network achieves 94.03\% accuracy, while the software-pruned and hardware-pruned networks achieve 92.21\% and 91.44\%.

Fig.~\ref{fig4}l presents layer-wise MAC precision for convolutional layers during training, showing minor BER variations due to device variations and potential RRAM cell failures. To mitigate these errors, two redundancy-aware correction mechanisms are implemented. The first reserves two of every 32 1T1R cells for fault tolerance. The second employs a backup memory region to replace faulty cells. Fig.~\ref{fig4}m left panel compares the operation (OPs) counts of convolutional layers during training. By employing dynamic kernel pruning, the computational workload of convolutional layers during training is reduced by 26.80\%. Fig.~\ref{fig4}m right panel compares the energy consumption of convolutional layers and fully connected (FC) layers during inference across different hardware platforms. The RRAM-based CIM system with pruning reduces energy consumption by 27.45\% compared to the system without pruning. Furthermore, compared to the NVIDIA GeForce RTX 4090, the pruned RRAM system achieves a 75.61\% reduction in energy consumption (normalized to the same technology node; see Supplementary Note 1 for details).
\begin{figure}[!t]
\centering
\includegraphics[width=0.9\linewidth]{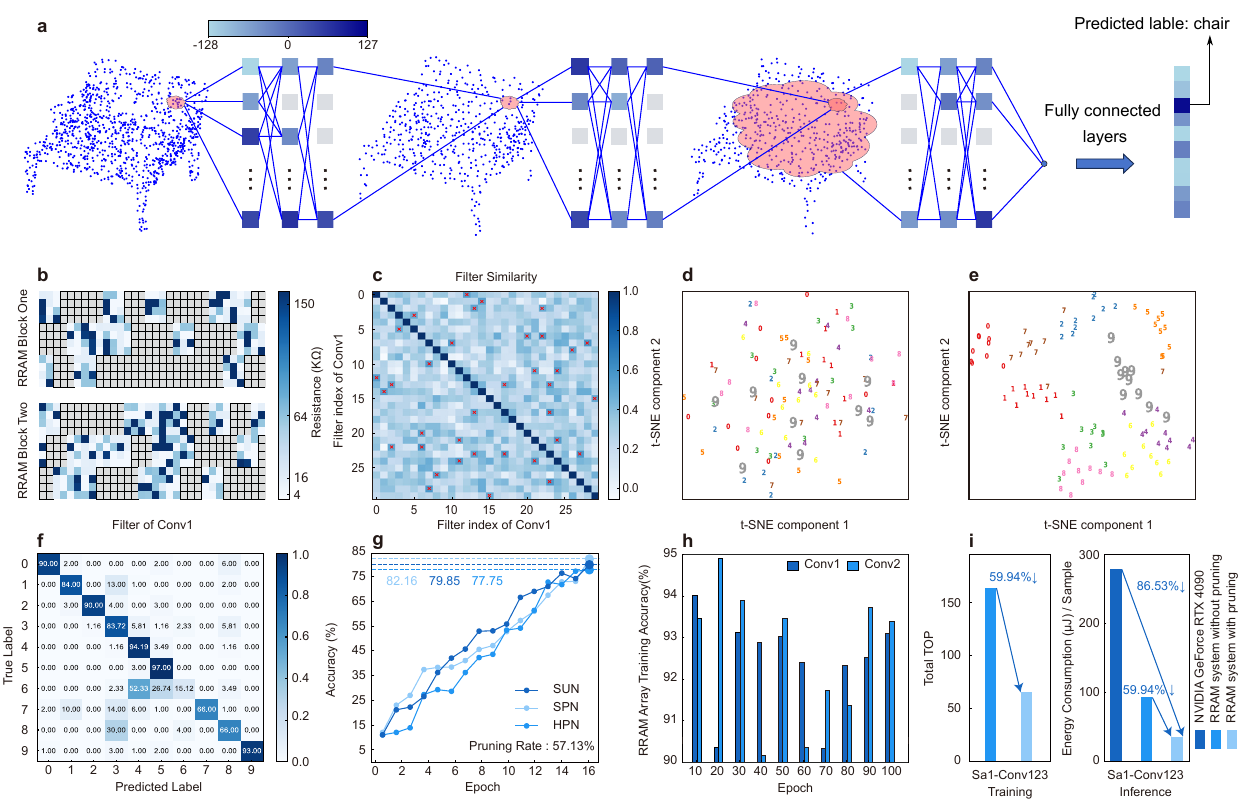}
\caption{\textbf{
Dynamic convolution filter pruning on PointNet++ for ModelNet10 classification.} 
\textbf{a,} Schematic illustration of PointNet++ forward pass, where hierarchical 1×1 convolutional layers extract geometric features before classification through fully connected layers.  
\textbf{b,} Mapping of 1×1 convolutional convolution filters of Conv1 on the RRAM array during the 10th training epoch.  
\textbf{c,} Convolution filters similarity matrix of Conv1, highlighting similarity using red crosses.  
\textbf{d-e,} t-SNE visualization of feature distribution before training and after training with dynamic convolution filter pruning.  
\textbf{f,} Normalized confusion matrix of classification results, where deeper colors indicate higher accuracy.  
\textbf{g,} Classification accuracy at a 57.13\% pruning rate for software-unpruned (SUN), software-pruned (SPN), and hardware-pruned (HPN) networks, achieving 79.85\%, 82.16\%, and 77.75\%, respectively.  
\textbf{h,} RRAM array MAC precision across training epochs.
\textbf{i,} Energy efficiency comparison between the NVIDIA GeForce RTX 4090 and the digital RRAM system with and without pruning.
}
\label{fig5}
\end{figure}

\section*{Dynamic PointNet Convolution Filter Pruning for ModelNet10 Classification}

This section describes the implementation and evaluation of dynamic convolution filter (hereafter referred to as “filter”) pruning in an RRAM-based PointNet++ neural network, which classifies 3D point clouds from the ModelNet10 dataset. Due to hardware constraints, only a subset of convolutional layers is deployed on-chip. We program each RRAM cell to 2-bit and use INT8 quantization for both weights and inputs, constraining their values within the range [-128, 127] (see Supplementary Figs. 2 and 3 for weight and quantization mapping, as well as convolution and similarity computation).

As shown in Fig.~\ref{fig5}a, the network follows the PointNet++ architecture. The input comprises three-dimensional point cloud data, represented by spatial coordinates (x, y, z) and initial feature vectors. Hierarchical 1×1 convolutional layers extract local geometric features. The final features are subsequently mapped to 10 output categories through two fully connected layers. Distinct filter weights are indicated using different colors, while gray represents pruned filters, removed dynamically based on the proposed pruning strategy. Fig.~\ref{fig5}b illustrates the mapping of a subset of first-layer 1×1 three-channel filters to resistive memory during the 10th training epoch. Each RRAM cell stores a 2-bit value, and after INT8 quantization, each weight is encoded using four RRAM cells. In epoch 10, 30 filters remain active (a subset of all filters). Their similarity matrix is shown in Fig.~\ref{fig5}c, where redundant filters are marked with red crosses. In this epoch, filters 13 and 23 are pruned, leaving 28 filters.

To assess the effectiveness of dynamic pruning, t-SNE dimensionality reduction is applied to 100 randomly selected feature vectors from the output of the classifier layer. Fig.~\ref{fig5}d visualizes feature distribution before training, whereas Fig.~\ref{fig5}e shows improved feature separation after training and dynamic filter pruning on the RRAM chip (see Supplementary Fig.5 about unpruning network feature separation). Classification performance after pruning is further evaluated using the normalized confusion matrix in Fig.~\ref{fig5}f, which measures classification accuracy across categories. Fig.~\ref{fig5}g examines the classification accuracy of software-unpruned (SUN), software-pruned (SPN), and hardware-pruned (HPN) networks at a 57.13\% pruning rate. The results indicate that SUN, SPN, and HPN achieve classification accuracies of 79.85\%, 82.16\%, and 77.75\%, respectively. The comparison between SUN and SPN demonstrates that dynamic filter pruning maintains high accuracy while significantly reducing computational cost. The similarity in accuracy between SPN and HPN further confirms that the RRAM-based CIM system performs comparably to software-based pruning on general-purpose hardware.

Fig.~\ref{fig5}h presents the MAC precision accuracy of three convolutional layers over 100 training epochs. Error correction mechanisms are implemented, reducing the BER to 0\%. Fig.~\ref{fig5}i (left) compares the OPs count of 1x1 convolutional layers during training. Dynamic filter pruning reduces the OPs count of convolutional layers by 59.94\% during training. Fig.~\ref{fig5}i (right) compares the energy consumption of convolutional layers during inference across different hardware platforms. With pruning, the RRAM-based CIM system reduces energy consumption by 59.94\% relative to its unpruned counterpart. Furthermore, compared to the NVIDIA GeForce RTX 4090, the pruned RRAM-based CIM system achieves an 86.53\% energy reduction (normalized to the same technology node; see Supplementary Note 1 for details).

\section*{Discussion}

This work demonstrates a fully digital CIM architecture based on reconfigurable RRAM arrays, integrated with a real-time weight pruning strategy. The system enables in-memory execution of both forward computation and similarity evaluation using bitwise logic operations, effectively eliminating data movement and ADC/DAC overhead. The architecture achieves energy-efficient inference and training by leveraging multi-bit non-volatile memory and logic reconfigurability. Compared to analog CIM and GPU platforms, it delivers up to energy reduction while maintaining zero bit error under operational conditions. These results confirm that logic-adaptive RRAM systems can support online structural sparsity in deep learning without accuracy degradation. While validated on compact convolutional and point cloud models, the scalability of this approach to larger architectures and more complex learning paradigms remains to be investigated. The current logic set supports binary operations; extending it toward more expressive or hierarchical primitives may further enhance task flexibility. These findings establish a foundation for next-generation edge AI hardware, offering a scalable and energy-efficient solution that bridges the gap between biological intelligence and artificial computing.

\section*{Methods}

\subsection*{Fabrication of the Reconfigurable Logic RRAM Chip}

The reconfigurable logic RRAM chip is fabricated using a standard 0.18$\mu$m CMOS process provided by United Microelectronics Corporation (UMC). Two \(512\times32\) 1T1R crossbar arrays are integrated, with resistive memory cells formed in the back-end-of-line (BEOL) between metal layers M5 and M6. Each cell consists of a top electrode (TE), bottom electrode (BE), and an active switching layer composed of a TiN/Ta\textsubscript{2}O\textsubscript{5}/TaO\textsubscript{x}/TiN stack. The full-chip layout is designed using Cadence Virtuoso. Peripheral digital control logic, including the shift-and-add and accumulator modules, is implemented separately on an external FPGA.

\subsection*{Device and Array Characterizations}
Electrical characterizations of both individual RRAM devices and the full array are performed at room temperature. Single-cell current–voltage (I–V) curves (Fig.~\ref{fig2}e,f) are measured using a Keysight B1500A semiconductor parameter analyzer under standard sweep conditions. Array-level evaluations (Fig.~\ref{fig2}e–i) are conducted using a custom-built testing system, based on the hardware framework described in the Supplementary Fig.1, with modifications to enable parallel array measurements.

\subsection*{Hardware System for Neural Network}
The hardware system comprises a 0.18$\mu$m RRAM chip, a printed circuit board (PCB), and an Xilinx ZCU102 FPGA. The RRAM chip operates in three modes: forming, programming, and computation. In forming mode, stochastic conductance states are initialized to generate random weights. In programming mode, resistive states are updated through set/reset pulses with write-verify to ensure accurate storage. Binary and 2-bit integer (INT2) values are distinguished using programmable voltage thresholds. During computation mode, the RRAM array performs vector–matrix multiplication and weight similarity evaluation entirely in memory. Logic operations such as AND and XOR are executed through reconfigurable units integrated in peripheral circuits. The FPGA manages vector input loading, result accumulation, and additional neural network functions such as activation and pooling.

\subsection*{Details of the Experimental Neural Networks}

For the MNIST classification task, a VGG16-based CNN with binarized weights is implemented to minimize energy consumption and latency. The network consists of three convolutional layers followed by a fully connected layer. The input is a 28×28 grayscale image. The first convolutional layer applies 32 binary 3×3 kernels (stride 1, padding 1), followed by ReLU activation and 2×2 max pooling. The second layer consists of 64 binary 3×3 kernels (stride 1, padding 1), with ReLU activation and 2×2 max pooling. The third layer uses 32 binary 3×3 kernels with ReLU activation. The extracted 32×7×7 feature maps are flattened into a 1568-dimensional vector, mapped to 10 output neurons in the final classification layer.

For ModelNet10 point cloud classification, a PointNet++ architecture processes 3D point cloud data represented by spatial coordinates $(x, y, z)$. The first Set Abstraction module (SA1) downsamples input points to 512, grouping 32 neighbors within a 0.2-radius and extracting features via an MLP (64, 64, 128). The second module (SA2) retains 512 points, expanding feature dimensions to 128, 128, and 256. The final module (SA3) aggregates all remaining points into a single 1024-dimensional global feature vector (MLP: 256, 512, 1024). Fully connected layers map the features through 512 and 256 dimensions before classification. Each fully connected layer is followed by batch normalization, ReLU activation, and dropout (0.5). The final classification layer outputs logits for 10 categories.

\section*{Acknowledgements}
This research is supported by the National Key R\&D Program of China (Grant No. 2023YFB2806300); Shenzhen Science and Technology Innovation Commission (Grant No. SGDX20220530111405040); Beijing Natural Science Foundation (Grant No. Z210006); and Hong Kong Research Grant Council (Grant Nos. 17205922, 17212923). This research is also partially supported by ACCESS – AI Chip Center for Emerging Smart Systems, sponsored by the Innovation and Technology Fund (ITF), Hong Kong SAR.

\section*{Competing Interests}
The authors declare no competing interests.

\bibliography{reference}

\end{document}


\maketitle
\newpage 
\section*{S1. Supplementary Figures}

\begin{figure}[!htbp]
\centering
\includegraphics[width=0.9\linewidth]{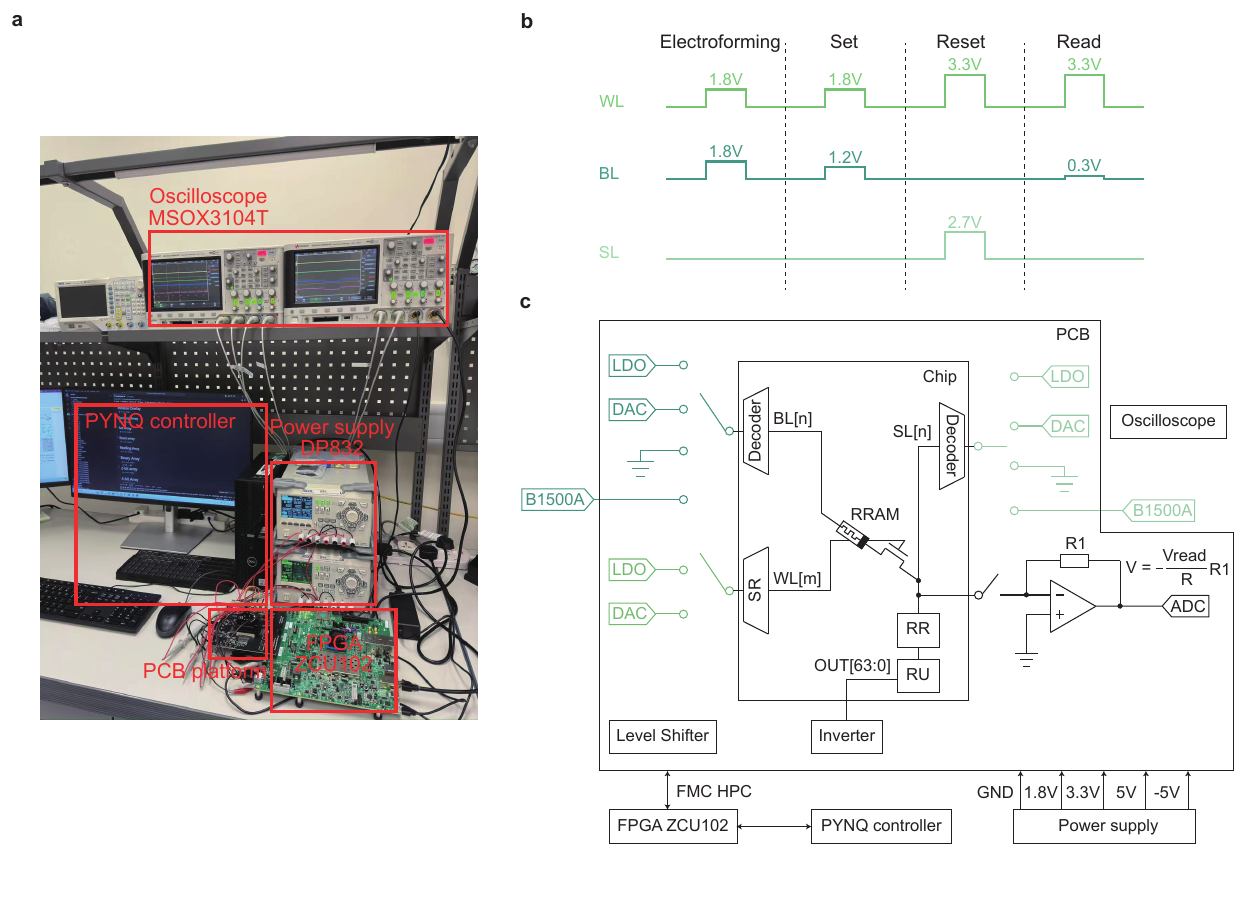}
\caption{\textbf{Experimental setup and RRAM operation scheme.}  
\textbf{a,} Photograph of the experimental setup, including two oscilloscopes (MSOX3104T) for signal monitoring, a PYNQ controller for digital control, an FPGA (ZCU102) for interfacing with a custom PCB platform, and two power supplies (DP832) for voltage regulation.  
\textbf{b,} Voltage waveform sequence for RRAM electroforming, set, reset, and read operations. The applied voltages on the word line (WL), bit line (BL), and source line (SL) are specified for each phase.  
\textbf{c,} Schematic of the programming system. The PCB integrates DACs, switches, and a low dropout regulator (LDO) to support different operational modes. During neural network training and inference, only the LDO and switches are required to program the RRAM resistance state, significantly reducing power consumption and hardware complexity. The resistance state is read through the Rref Read and Reconfigurable Unit modules integrated within the chip, enabling efficient and low-overhead resistance state retrieval. For experimental studies of RRAM array performance, programming and characterization can be performed using on-board DACs and switches for fast and coarse tuning. During read operations, a voltage is applied to the BL, and the resulting current is converted into a voltage via a transimpedance amplifier (TIA) before being digitized by the ADC to estimate the RRAM resistance state. For precise programming and high-accuracy resistance measurement, an external B1500A unit provides fine-tuned voltage control.
}
\label{figs1}
\end{figure}
\clearpage

\begin{figure}[!htbp]
\centering
\includegraphics[width=0.9\linewidth]{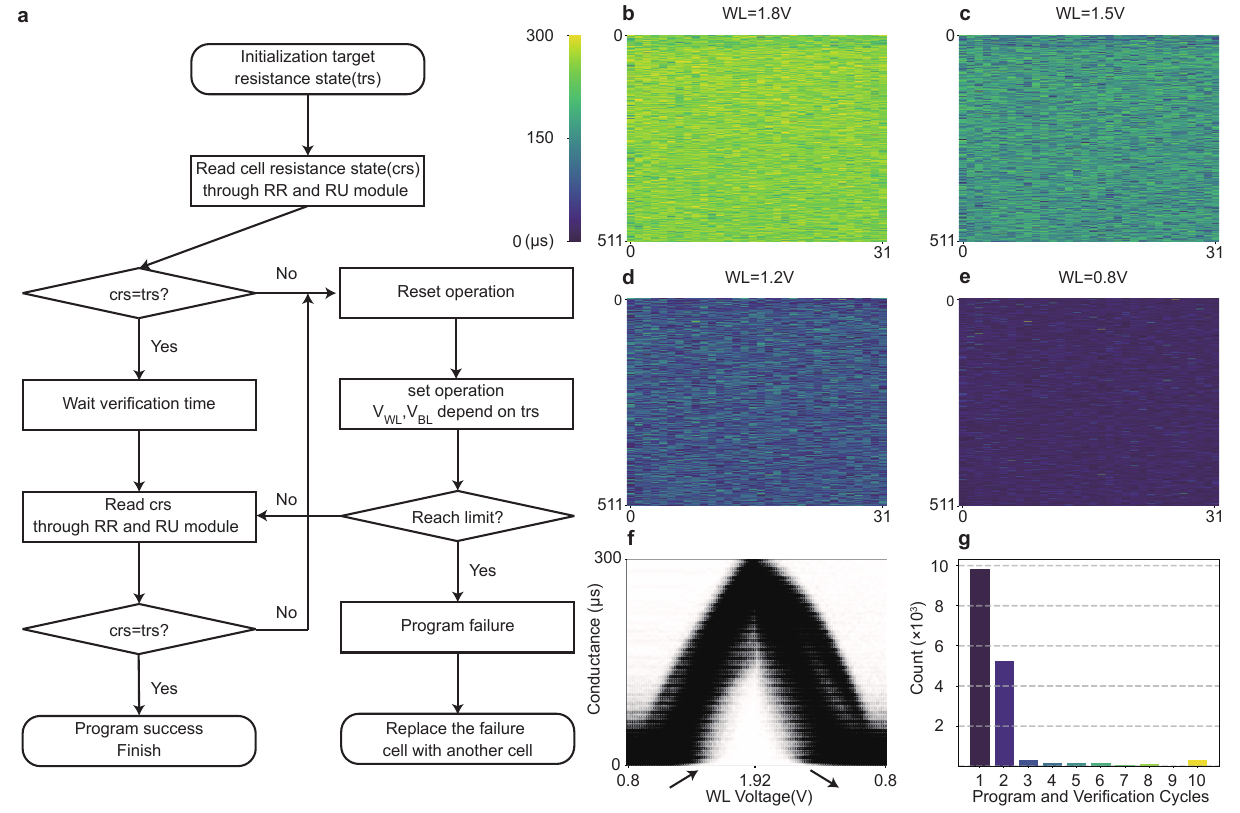}
\caption{\textbf{RRAM programming flowchart and characteristics.} 
\textbf{a}, Programming and verification flowchart for tuning each RRAM cell to a target resistance state (trs). The process begins by reading the current resistance state (crs) through the Read Rref (RR) module and the Reconfigurable Unit (RU) module and comparing it with the trs. If the states do not match, a reset or set operation is applied depending on the deviation. This loop of read and verification continues until the target state is achieved or the maximum allowed number of cycles is reached, in which case the cell is marked as a failure and replaced. 
\textbf{b--e}, Conductance mapping of the 512 × 32 RRAM array under four different word-line (WL) voltages during programming to the INT2 state: \textbf{b}, 1.8 V; \textbf{c}, 1.5 V; \textbf{d}, 1.2 V; \textbf{e}, 0.8 V.
\textbf{f}, Global conductance distribution of the array as a function of WL voltage. The WL voltage is swept from 0.8 V to 1.92 V in 0.02 V increments and then decreased back to 0.8 V with the same step size.
\textbf{g}, Histogram showing the number of program-and-verify cycles required for all cells in the array to reach the target resistance state. The majority of cells reach the target within 1--2 cycles.
}
\label{figs2}
\end{figure}
\clearpage

\begin{figure}[!htbp]
\centering
\includegraphics[width=0.9\linewidth]{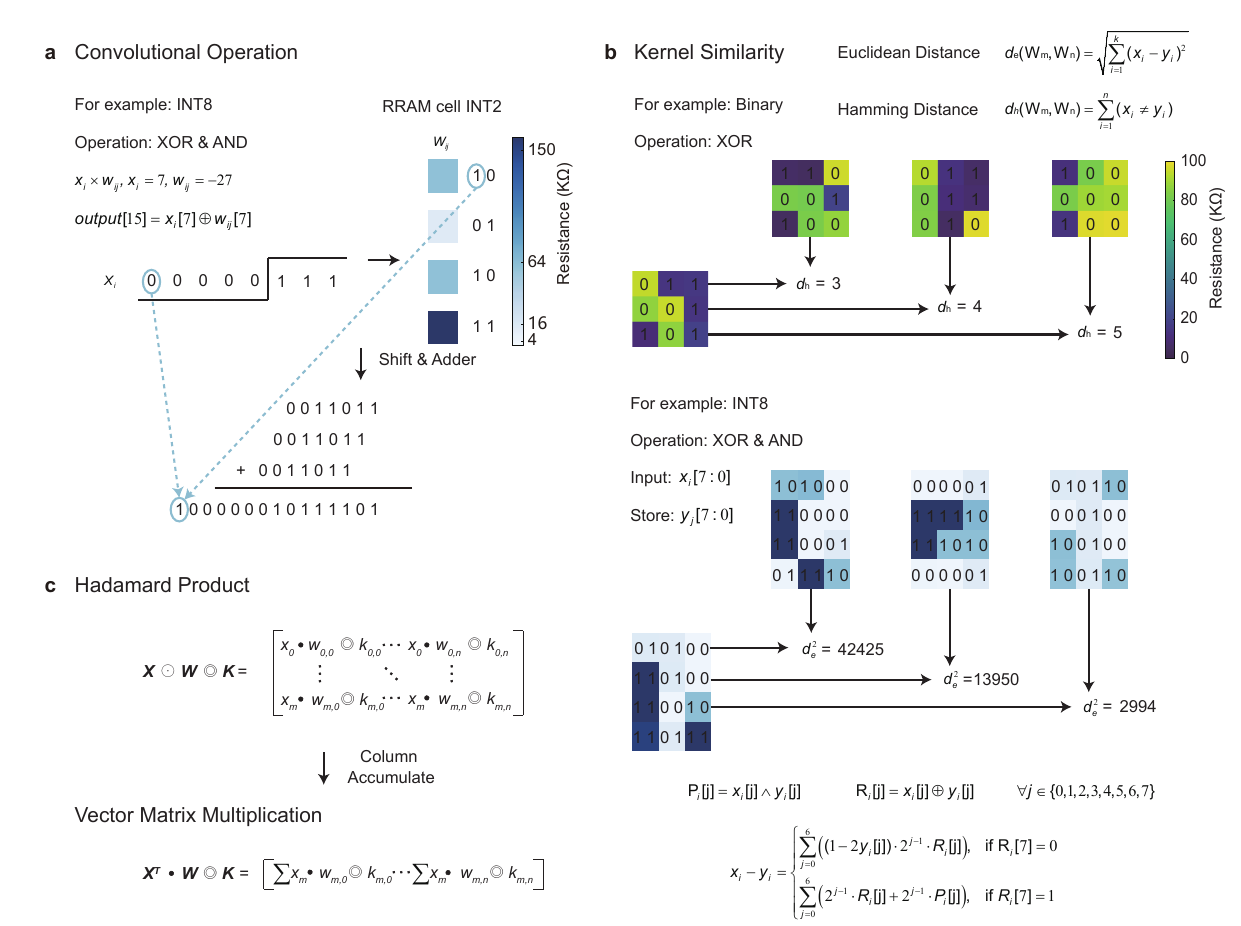}
\caption{\textbf{RRAM-based convolution operation and weight similarity.} 
\textbf{a,} Convolution operation using signed 8-bit integer (INT8) inputs and weights. Each 2-bit segment of the weight is stored in a single RRAM cell, with resistance values representing the encoded bits. Computation is performed using bitwise XOR and AND operations. For example, given $x_i = 7$ (00000111) and $w_{ij} = -27$ (10011011), the output sign bit is computed as $\text{output}[15] = x_i[7] \oplus w_{ij}[7]$. The partial products are accumulated through a shift-and-adder group to generate the final output.  
\textbf{b,} Weight similarity evaluation based on either Hamming or Euclidean distance. For binary weights, the Hamming distance $d_h$, computed via XOR, can approximate the Euclidean distance $d_e$. For INT8 weights, XOR and AND operations are applied bitwise between two weight vectors $x_i$ and $y_i$ to compute $x_i - y_i$. The squared Euclidean distance $d_e^2$ is then derived from the bitwise logic values $P_j = x_j \land y_j$ and $R_j = x_j \oplus y_j$ using the provided formula.  
\textbf{c,} Ternary computation based on the Hadamard product and accumulation. The element-wise operation $X \odot W \circledcirc K$ is computed, followed by column-wise accumulation to yield the result of ternary vector-matrix multiplication $X^\top \cdot W \circledcirc K$, where $\circledcirc$ represents logical operations such as NAND, AND, XOR, or OR.}
\label{figs3}
\end{figure}
\clearpage

\begin{figure}[!htbp]
\centering
\includegraphics[width=0.9\linewidth]{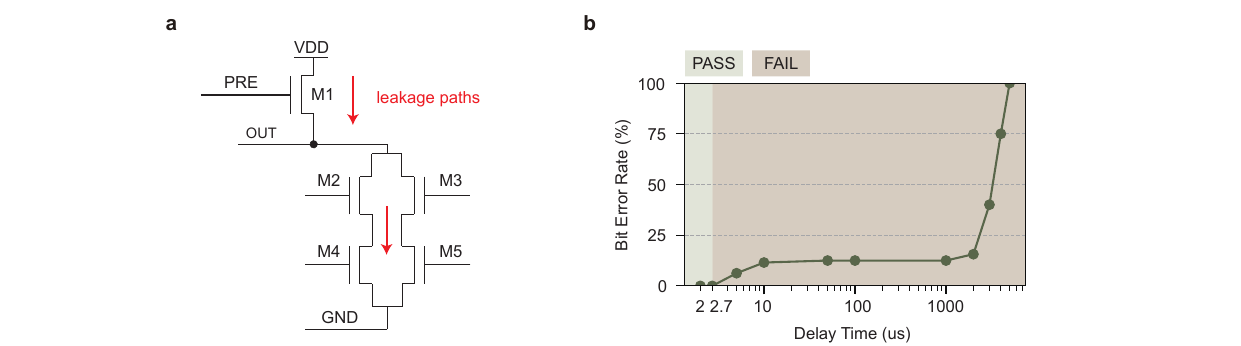}
\caption{\textbf{Impact of leakage on compute accuracy.}
\textbf{a,} Schematic illustration of leakage paths~\cite{yan20221} in the precharge-compute structure. During the precharge phase, the signal line OUT is pulled up to 1.8V through transistor M1. However, during the subsequent compute phase, if the operation time is prolonged, charge on OUT can dissipate through leakage paths formed by transistors M2-M5, leading to a voltage drop and resulting in incorrect computation outcomes. 
\textbf{b,} Measured relationship between compute delay time and bit error rate (BER). When the delay exceeds a critical threshold ($\sim$2.7\,$\mu$s), leakage-induced charge loss leads to a rapid increase in BER.
}
\label{figs4}
\end{figure}
\clearpage

\begin{figure}[!htbp]
\centering
\includegraphics[width=0.9\linewidth]{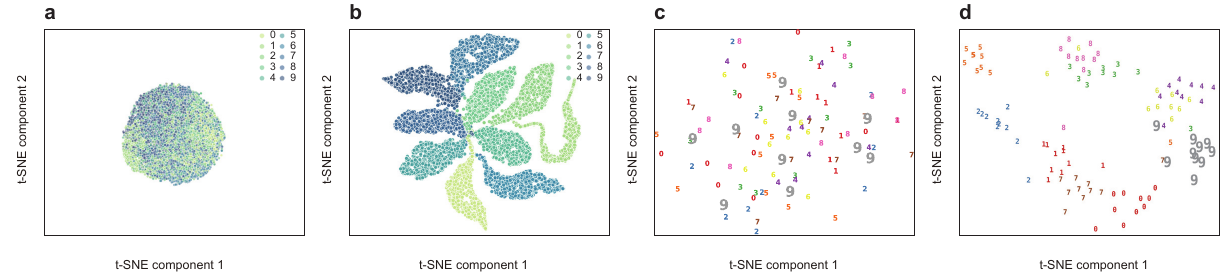}
\caption{\textbf{t-SNE visualization of unpruned networks.}
\textbf{a,} Feature distribution before training in a CNN-based network, showing high inter-class overlap. 
\textbf{b,} Feature distribution after training in the same CNN, with improved clustering and separability. 
\textbf{c,} Feature distribution before training in a PointNet++ network for point cloud classification. 
\textbf{d,} Feature distribution after training in PointNet++, exhibiting clear inter-class separation. 
}
\label{figs5}
\end{figure}
\clearpage

\section*{S2. Supplementary Tables}

\begin{table}[ht]
\centering
\caption{Area and energy consumption breakdown by module.}
\begin{tabular}{c|c|c|c|c|c|c|c}
\toprule
 & \textbf{WRC} & \textbf{BSIC} & \textbf{RRAM} & \textbf{RR} & \textbf{RU} & \textbf{S \& A} & \textbf{ACC} \\
\midrule
\makecell[c]{\textbf{Area} \\ (mm$^2$)} & $0.106$ & $0.028$ & $0.534$ & $0.005$ & $0.001$ & $0.037$ & $0.155$ \\
\midrule
\makecell[c]{\textbf{Energy} \\ (pJ)} & $2690.400$ & $68.682$ & $0.576$ & $34.001$ & $22.056$ & $269.12$ & $907.136$  \\
\bottomrule
\end{tabular}
\label{tab:area_energy}
\end{table}

\vspace{1em}
\noindent
\textbf{Abbreviations:}
WRC – WL Driver \& RU Controller; 
BSIC – BL/SL Driver Circuits \& Input Controller; 
RRAM – Resistive Random-Access Memory Array; 
RR – Rref Read module; 
RU – Reconfigurable Unit; 
S \& A – Shift and Adder Group; 
ACC – Accumulator.

\vspace{1em}
\noindent
\textbf{Note.} The reported energy consumption values correspond to the operation involving parallel 64-bit AND computations across the system, including the subsequent shift and accumulation processes required for final result generation.

\clearpage

\begin{table}[ht]
\centering
\caption{Architecture of the VGG16-based model (Task 1) and PointNet++ model (Task 2).}
\begin{tabular}{c|p{6.4cm}|p{7.2cm}}
\toprule
\textbf{Layer} & \textbf{VGG16-based model} & \textbf{PointNet++ model} \\
\midrule
1 & \ttfamily BinaryConv2d(1, 32, kernel\_size=3, padding=1) $\rightarrow$ ReLU $\rightarrow$ MaxPool2d(2) 
  & \ttfamily SA1: in\_channels=$C_{in}$, mlp=[64, 64, 128], group\_all=False \\
\midrule
2 & \ttfamily BinaryConv2d(32, 64, kernel\_size=3, padding=1) $\rightarrow$ ReLU $\rightarrow$ MaxPool2d(2)
  & \ttfamily SA2: in\_channels=131, mlp=[128, 128, 256], group\_all=False \\
\midrule
3 & \ttfamily BinaryConv2d(64, 32, kernel\_size=3, padding=1) $\rightarrow$ ReLU
  & \ttfamily SA3: in\_channels=259, mlp=[256, 512, 1024], group\_all=True \\
\midrule
4 & \ttfamily Flatten to vector of size 32 $\times$ 7 $\times$ 7
  & \ttfamily Flatten to vector of size 1024 \\
\midrule
5 & \ttfamily Linear(1568, 10) for classification
  & \ttfamily Linear(1024, 512, bias=False) $\rightarrow$ BN(512) $\rightarrow$ ReLU $\rightarrow$ Dropout(0.5) \\
\midrule
6 & --
  & \ttfamily Linear(512, 256, bias=False) $\rightarrow$ BN(256) $\rightarrow$ ReLU $\rightarrow$ Dropout(0.5) \\
\midrule
7 & --
  & \ttfamily Linear(256, num\_classes) for classification \\
\bottomrule
\end{tabular}
\label{tab:model_architecture}
\end{table}

\noindent\textbf{Note.} The complete implementation of both models is available at: \\
\url{https://github.com/wangsongq/Dynamic-Kernel-Pruning.git}

\clearpage

\section*{S3. Supplementary Notes}

\subsection*{Note1. Computational Cost Estimation}

The term "Ops." refers to the number of multiply-accumulate operations (MACs) required during a single forward inference. For different layer types, the computations are estimated as follows:

\subsubsection*{Convolutional Layer}
\begin{equation}
\text{Ops.} = 2 \times C_{\text{in}} \times C_{\text{out}} \times k_h \times k_w \times H_{\text{out}} \times W_{\text{out}}
\label{eq:Conv},
\end{equation}
where $C_{\text{in}}$ and $C_{\text{out}}$ denote the number of input and output channels, $k_h$ and $k_w$ denote the kernel height and width, and $H_{\text{out}}, W_{\text{out}}$ denote the output feature map dimensions. The factor 2 accounts for both multiplication and accumulation.

\subsubsection*{Fully Connected Layer}
\begin{equation}
\text{Ops.} = 2 \times w_h \times w_w
\label{eq:FC},
\end{equation}
where $w_h$ and $w_w$ denote the height and width of the weight matrix in the linear layer.

\subsubsection*{NVIDIA GeForce RTX 4090}

To establish a quantitative energy-efficiency baseline for conventional digital accelerators, we estimate the per-operation and per-bit energy consumption of the NVIDIA GeForce RTX 4090 when executing INT8 operations. According to publicly available data from \href{https://llm-tracker.info/GPU-Comparison}{LLM Tracker} and \href{https://www.techpowerup.com/gpu-specs/geforce-rtx-4090.c3889}{TechPowerUp}, the peak INT8 tensor compute performance of the RTX 4090 reaches 660.6~TOPS (dense), with a power draw of 450~W under full load. The resulting energy cost per INT8 operation is given by:

\begin{equation}
\text{Energy per Operation} = \frac{\text{Power Draw}}{\text{Peak INT8 Performance}} = \frac{450~\text{W}}{660.6 \times 10^{12}~\text{Ops/s}} = 0.6812~\text{pJ/op}
\label{eq:4090_INT8}
\end{equation}

\subsubsection*{Our system}

To evaluate the energy efficiency of our proposed system in a realistic serial multiplication architecture, we model the full execution of a multiplication and accumulation across 32 parallel column units. 

The overall power consumption of the system, accounting for WL/SR driver (134.52~mW), bit-line decoder (3.4341~mW), RRAM array (0.0288~mW), read units (1.6992~mW), reconfigurable units (1.104~mW), shift-and-adder units (13.456~mW), and accumulators (45.3568~mW), amounts to 199.60~mW. The average energy consumption per operation is thus:

\begin{equation}
\text{Energy per operation} = \frac{199.60\ \text{mW} \times 22.5\ \text{ns}}{64} = 70.17\ \text{pJ/OP}
\end{equation}

To provide a normalized comparison with modern digital platforms, we scale the energy consumption based on voltage and frequency using standard dynamic power scaling laws. Specifically, assuming a 0.8~V supply and a clock frequency of 1.8~GHz (typical for modern GPUs), we apply:

\begin{equation}
\text{Scaled Energy} = 70.17~\text{pJ} \cdot \left(\frac{0.8}{3.3}\right)^2 \cdot \left(\frac{100~\text{MHz}}{1.8~\text{GHz}}\right) \approx 0.229~\text{pJ/OP} = 0.229~\text{pJ/OP}
\end{equation}

Note that this estimation excludes process technology scaling (e.g., from 180~nm to 7~nm) and is thus conservative.

\clearpage



\bibliographystyle{unsrt}
\bibliography{reference}